\documentclass[aps,prd,twocolumn,10pt,superscriptaddress,nofootinbib,
preprintnumbers,amsmath,amssymb,floatfix,showkeys]{revtex4-2}

\usepackage[utf8]{inputenc}   
\usepackage[T1]{fontenc}

\usepackage[parfill]{parskip}
\usepackage{graphicx}
\usepackage{xcolor}           
\usepackage{float}
\usepackage{adjustbox}
\usepackage{bm}
\usepackage{hyperref}         
\usepackage{orcidlink}        

\begin{document}

\title{A Closed-Form Master Solution for Gravitational Decoupling of a Unified Family of Isotropic Stellar Seeds}

\author{Anirudh Pradhan \orcidlink{0000-0002-1932-8431}}
\email{pradhan.anirudh@gmail.com}
\affiliation{Centre for Cosmology, Astrophysics and Space Science,
GLA University, Mathura-281\,406, Uttar Pradesh, India}

\author{Safiqul Islam 
\orcidlink{0000-0003-1373-4137}
}
\thanks{Corresponding author}
\email{sislam@kfu.edu.sa}
\affiliation{Department of Mathematics and Statistics, College of Science,
King Faisal University, P.O. Box 400, Al Ahsa 31982, Saudi Arabia}

\author{Ajit Kumar
\orcidlink{0000-0002-3015-3185}
}
\email[]{ajit.chauhan79@gmail.com}
\affiliation{ Department of Mathematics, Faculty of Engineering,
 Teerthanker Mahaveer University, 
 Moradabad-244001, India }

\author{Muhammad Aamir \orcidlink{0000-0002-5782-0064}
}
\email[]{msadiq@kfu.edu.sa}
\affiliation{Department of Physics, College of Science,\\ King Faisal University, P.O. Box 400, Al Ahsa 31982, Saudi Arabia}

\begin{abstract}
We present a unification scheme for anisotropic compact stars generated via minimal geometric deformation (MGD), built on a two-parameter seed family $\nu(r)=\text{Ln}[C(1+ar^{2n})^{m}]$ whose radial metric potential $\mu(r)=e^{-\lambda(r)}$ is fixed in closed form by the isotropy condition. At $n=1$ the family reproduces Tolman IV, Korkina--Orlyanskii, Heintzmann IIa, and Durgapal IV--V as the particular cases $m=1,2,3,4,5$. The key result is a single closure of the MGD $\theta$-sector, through one source function $g(r)$, that integrates the decoupler equation in closed form, in terms of the Gauss hypergeometric function ${}_2F_1$, for \emph{arbitrary} $(n,m)$: the anisotropic extension of the $n=1$ family is thus obtained from one master formula. We derive the effective anisotropic matter content, the Darmois--Israel matching to the exterior Schwarzschild vacuum, and the full set of physical acceptability criteria constraining the parameter space, and map the admissible region in compactness and decoupling strength explicitly for the case $n=1$: Tolman IV and Korkina--Orlyanskii admit physically consistent anisotropic deformations at arbitrarily high compactness, while Heintzmann IIa and Durgapal IV--V are bounded by a compactness ceiling beyond which no value of the decoupling parameter restores causality. Using the same closed-form matter content, we compute the quadrupolar tidal Love number and dimensionless tidal deformability across the admissible region, finding that the decoupling systematically enhances both relative to the isotropic seed at fixed compactness, with a sharp rise near the causal ceiling in the Heintzmann and Durgapal branches. 
\end{abstract}


\maketitle


\section{Introduction}\label{sec1}

Relativistic compact objects, white dwarfs, neutron stars, and the more speculative exotic stars, are unique laboratories to test general relativity (GR) and the behavior of matter under extreme conditions of density and pressure. Since Schwarzschild's interior solution for an incompressible, locally isotropic fluid sphere \cite{weinberg}, the search for exact, physically viable solutions of the Einstein field equations describing static, spherically symmetric stellar configurations has remained a central problem of relativistic astrophysics. The isotropic assumption $p_{r}=p_{\perp}\equiv \bar p$, while mathematically convenient, is a simplification: for matter densities beyond nuclear saturation, several independent physical mechanisms are expected to render the local pressure anisotropic, $p_{r}\neq p_{\perp}$. Among these, the existence of a solid crust or a superfluid core, pion condensation, the presence of type-3A superfluids, strong magnetic fields, slow rotation, phase transitions, and the intrinsic anisotropy of the equation of state (EoS) itself at supranuclear densities have all been invoked in the literature \cite{RUDERMAN,BOWERS,herre}. Bowers and Liang \cite{BOWERS} first showed systematically that local anisotropy can modify the maximum equilibrium mass, the surface redshift, and the stability of a compact star in a non-negligible way, a result later confirmed and extended by many authors using both physically motivated and phenomenological anisotropy profiles \cite{GOKHROO,MAK,IVANOV,LEMA}.

Building anisotropic interior solutions directly, by postulating a functional form for $\Delta \equiv p_{\perp}-p_{r}$ together with one of the metric potentials, is a well-trodden scheme. In this regard, a systematic alternative was provided by the gravitational decoupling (GD) technique in its minimal geometric deformation (MGD) realization \cite{O1,O2}. In this approach, one starts from a known and well-behaved isotropic seed solution $\{\bar\rho,\bar p,\xi,\mu\}$ of GR sourced by $\bar T_{\mu\nu}$, and extends the total energy-momentum content according to $\bar T_{\mu\nu}\to T_{\mu\nu}=\bar T_{\mu\nu}+\alpha\,\theta_{\mu\nu}$, where $\theta_{\mu\nu}$ is an additional gravitational source, minimally coupled to the seed through the dimensionless parameter $\alpha$. The radial metric potential alone is deformed, $e^{-\lambda}\mapsto \mu(r)+\alpha f(r)$, which has the remarkable effect of splitting the highly non-linear Einstein system into two independently conserved sectors: the original seed equations, already solved, and a new, purely gravitational $\theta$-sector that can be closed and solved on top of any admissible seed. This ``divide and conquer'' strategy has proven extraordinarily fruitful, having been used to build anisotropic stars \cite{O1,O2,P1,E5,P2,P3,E1,E2,E3,E4,M1,Tello-Ortiz:2020euy,Tello-Ortiz:2020nuc,Tello-Ortiz:2020ydf,Tello-Ortiz:2023poi,Tello-Ortiz:2025eyj,Azmat:2022fad,Maurya:2019hds,Maurya:2019sfm,Maurya:2019xcx,Maurya:2020djz,Maurya:2020ebd,Maurya:2021aio,Maurya:2021fuy,Maurya:2021hty,Maurya:2021tca,Morales:2018nmq,Morales:2018urp,Estrada:2018zbh}, to study the interior structure of dark, wormhole, and braneworld-inspired compact objects \cite{D1,D2,D3,D4}, and to explore extensions of GR and dark-sector couplings.

Within this program, the $\theta$-sector -- being undetermined by the field equations alone -- must always be closed with additional, physically motivated input. Tolman IV, Korkina--Orlyanskii, Heintzmann IIa, and the Durgapal IV and V solutions -- five of the handful of isotropic solutions singled out as ``well behaved'' in the classical survey of Delgaty and Lake \cite{delgaty} -- have each been anisotropized via MGD in a separate publication, with a closure of the $\theta$-sector chosen for that particular seed, whether through a mimic constraint, a Weyl-fluid profile, or a regularity condition on the anisotropy factor. Each such closure is a legitimate and well-motivated choice on its own terms; what none of them was designed to do is extend, as it stands, to a different seed without being rederived. A parallel, non-MGD route to the same set of solutions was opened by Thirukkanesh, Ragel, Sharma and Das \cite{thiru}, who showed that all five metrics are recovered from a single generating potential, and who used a direct anisotropic-generating algorithm, rather than GD, to produce closed-form anisotropic extensions of the whole family at once. What is still missing is the MGD counterpart of that unification: a single, closed-form decoupler function $f(r)$, valid for arbitrary $(n,m)$, that anisotropizes the whole seed family under one common prescription.

This is precisely the gap we fill in this work. We take as seed the two-parameter family generated by
\begin{equation}
\nu(r)=\text{Ln}\left[C\left(1+ar^{2n}\right)^{m}\right],
\end{equation}
solve the isotropy condition to obtain the associated $\mu(r)$ in closed form (in terms of the Appell hypergeometric function $F_{1}$), and identify the values of $(n,m)$ that reproduce each of the five classical solutions quoted above. We then close the $\theta$-sector by prescribing the source function $g(r)\equiv\theta^{r}_{r}-\theta^{\varphi}_{\varphi}$ in a form specifically engineered so that the linear, first-order ordinary differential equation obeyed by $f(r)$ can be integrated in terms of elementary and Gauss hypergeometric functions for {any} $(n,m)$, rather than only for a hand-picked value. Physically, this means that the anisotropy induced by MGD is fixed, for the whole $n=1$ family at once, by a single functional prescription -- as opposed to five independent, unrelated prescriptions scattered across the literature. Beyond its economy, this construction provides a natural and rigorous way to compare, within a common mathematical scheme, how the anisotropic imprint of $\theta_{\mu\nu}$ depends on the seed's compactness profile (encoded in $m$) and on its radial ``sharpness'' (encoded in $n$).

The remainder of this paper is organized as follows. In Sec.~\ref{sec2} we review the MGD formalism in the form required for our construction. In Sec.~\ref{sec3} we present the unified seed, its known special cases, and the closed-form solution of the $\theta$-sector. Section~\ref{sec4} works out the effective anisotropic matter content and the matching to the exterior vacuum. Section~\ref{sec5} collects the physical acceptability criteria that constrain the parameter space $(a,A,B,C,D,\alpha,n,m)$, and Sec.~\ref{sec6} discusses the astrophysical application of the model, including its relevance for the tidal deformability of coalescing compact stars. We conclude in Sec.~\ref{sec7}. Throughout, we work in relativistic geometrized units $c=G=1$ and adopt the mostly negative signature $\{+,-,-,-\}$.

\section{The Gravitational Decoupling Method}\label{sec2}

One of the central open problems in constructing spherically symmetric, static stellar configurations within GR is how to consistently introduce local pressure anisotropy into the matter distribution while retaining full control over the resulting geometry. In principle one could postulate from the outset an anisotropic energy-momentum tensor and solve the field equations directly. However, when the entire matter content is lumped into a single, generic imperfect fluid, it becomes impossible to disentangle which physical ingredient is actually responsible for $p_{r}\neq p_{\perp}$; one can only state that the fluid \emph{is} anisotropic, without any handle on the underlying mechanism, nor any controlled way to switch the anisotropy off and recover a specific, known isotropic limit. This is the practical difficulty that the gravitational GD, in its MGD realization \cite{O1}, was designed to overcome.

The starting point is an isotropic matter distribution,
\begin{equation}\label{eq1}
\bar{T}_{\mu\nu}=\left(\bar{\rho}+\bar{p}\right)\chi_{\mu}\chi_{\nu}-\bar{p}\,g_{\mu\nu},
\end{equation}
with $\bar\rho$ and $\bar p$ the isotropic density and pressure (barred quantities denote the isotropic seed throughout this work). The source is then extended by a new sector $\theta_{\mu\nu}$, minimally coupled through a dimensionless constant $\alpha$,
\begin{equation}\label{eq2}
\bar{T}_{\mu\nu}\;\longrightarrow\; T_{\mu\nu}=\bar{T}_{\mu\nu}+\alpha\,\theta_{\mu\nu}.
\end{equation}
The nature of $\theta_{\mu\nu}$ is left unspecified \emph{a priori}: depending on the physical context it can encode an additional fluid, a scalar or vector field, a braneworld correction, or even the dark sector of the Universe \cite{O2}. In canonical (Schwarzschild-like) coordinates, the line element for a static, spherically symmetric configuration reads
\begin{equation}\label{eq3}
ds^{2}=e^{\nu(r)}dt^{2}-e^{\lambda(r)}dr^{2}-r^{2}d\Omega^{2},
\end{equation}
with $\chi^{\mu}=e^{-\nu/2}\delta^{\mu}_{t}$ the static 4-velocity. Substituting Eqs.~\eqref{eq2}--\eqref{eq3} into the Einstein equations $G_{\mu\nu}=-8\pi T_{\mu\nu}$ gives\footnote{Primes denote differentiation with respect to $r$ throughout.}
\begin{eqnarray}\label{eq12}
{{\rm -e}^{-\lambda}}\left( {\frac {\lambda^{{\prime}}}{r}}-\frac{1}{r^2}\right)-\frac{1}{r^2}&=&-8\pi\left(\bar{\rho}+\alpha\theta^{t}_{t}\right), \\ \label{eq13}
{{\rm e}^{-\lambda}} \left( {\frac {\nu^{{\prime}}}{r}}+\frac{1}{r^2}\right) -\frac{1}{r^2}&=&8\pi\left(\bar{p}-\alpha\theta^{r}_{r}\right),\\ \label{eq14}
\frac{{\rm e}^{-\lambda}}{4} \left(2 \nu^{{\prime\prime}}+\nu^{\prime 2}+2{\frac {\nu^{{\prime}}-\lambda^{{\prime}}}{r}}-\nu^{\prime}\lambda^{\prime} \right) &=&8\pi\left(\bar{p}-\alpha\theta^{\varphi}_{\varphi}\right),
\end{eqnarray}
together with the total conservation equation
\begin{equation}\label{eq15}
\bar p' +\frac{\nu^{\prime}}{2} (\bar{\rho}+\bar{p})+\alpha\left[ \frac{\nu^{\prime}}{2}\left(\theta^{t}_{t}-\theta^{r}_{r}\right)-\left(\theta^{r}_{r}\right)'-\frac{2}{r}\,(\theta^{\varphi}_{\varphi}-\theta^{r}_{r})\right]=0.
\end{equation}
The effective (total) density and pressures are then
\begin{eqnarray}\label{eq16}
\rho&\equiv& \bar{\rho}+\alpha\theta^{t}_{t}, \\ \label{eq17}
p_{r}&\equiv&\bar{p}-\alpha\theta^{r}_{r}, \\ \label{eq18}
p_{t}&\equiv&\bar{p}-\alpha\theta^{\varphi}_{\varphi},
\end{eqnarray}
so that the source is genuinely anisotropic, $p_r\neq p_t$, whenever $\theta^{r}_{r}\neq\theta^{\varphi}_{\varphi}$. At first sight, coupling in $\theta_{\mu\nu}$ only makes the problem harder: the number of unknowns grows while the field equations remain three. The key observation of Ovalle \cite{O1} is that this is not the case if one performs the following geometric transformation on the metric potentials,
\begin{equation}\label{17}
\nu(r) \mapsto \xi(r)+\alpha h(r), \qquad e^{-\lambda(r)} \mapsto \mu(r) +\alpha f(r),
\end{equation}
with $h(r)$ and $f(r)$ the \emph{deformation} (or \emph{decoupler}) functions of the temporal and radial metric components, respectively. The \emph{minimal} geometric deformation corresponds to deforming only the radial metric potential, $h=0$, $f\neq0$, leaving $\nu(r)=\xi(r)$ untouched. Substituting Eq.~\eqref{17} with $h=0$ into Eqs.~\eqref{eq12}--\eqref{eq14}, the system splits \emph{exactly} into two independent subsystems. The first,
\begin{eqnarray}\label{eq23}
8\pi\bar{\rho}&=&\frac{1}{r^{2}}-\frac{\mu}{r^{2}}-\frac{\mu^{\prime}}{r},\\\label{eq24}
8\pi \bar{p}&=&-\frac{1}{r^{2}}+\mu\left(\frac{1}{r^{2}}+\frac{\xi^{\prime}}{r}\right),\\\label{eq25}
8\pi \bar{p}&=&\frac{\mu}{4}\left(2\xi^{\prime\prime}+\xi^{\prime2}+2\frac{\xi^{\prime}}{r}\right)+\frac{\mu^{\prime}}{4}\left(\xi^{\prime}+\frac{2}{r}\right),
\end{eqnarray}
is nothing but the ordinary Einstein system for a perfect fluid with metric potentials $\{\xi,\mu\}$, obeying the familiar hydrostatic equilibrium equation
\begin{equation}\label{eq26}
\bar{p}^{\prime}+\frac{\xi^{\prime}}{2}\left(\bar{\rho} +\bar{p}\right)=0.
\end{equation}
This is the \emph{seed} system: any known, well-behaved isotropic solution $\{\bar\rho,\bar p,\xi,\mu\}$ of GR solves it identically. The second subsystem,
\begin{eqnarray}\label{eq27}
8\pi\theta^{t}_{t}&=&-\frac{f}{r^{2}}-\frac{f^{\prime}}{r},
\\ \label{eq28}
8\pi\theta^{r}_{r}&=&-f\left(\frac{1}{r^{2}}+\frac{\xi^{\prime}}{r}\right),  \\  \label{eq29}
8\pi\theta^{\varphi}_{\varphi}&=&-\frac{f}{4}\left(2\xi^{\prime\prime}+\xi^{\prime2}+2\frac{\xi^{\prime}}{r}\right)-\frac{f^{\prime}}{4}\left(\xi^{\prime}+\frac{2}{r}\right),
\end{eqnarray}
often referred to as \emph{quasi-Einstein} equations because of the missing centrifugal term $1/r^{2}$, governs the new gravitational source $\theta_{\mu\nu}$ \emph{on the already fixed background} $\xi(r)$. Crucially, Eqs.~\eqref{eq27}--\eqref{eq29} obey their own, separate conservation law,
\begin{equation}\label{eq30}
\left(\theta^{r}_{r}\right)^{\prime}-\frac{\xi^{\prime}}{2}\left(\theta^{t}_{t}-\theta^{r}_{r}\right)-\frac{2}{r}\left(\theta^{\varphi}_{\varphi}-\theta^{r}_{r}\right)=0,
\end{equation}
which is a linear combination of Eqs.~\eqref{eq27}--\eqref{eq29} and not an independent equation. The simultaneous validity of Eqs.~\eqref{eq26} and \eqref{eq30} is the mathematical statement that the seed source $\bar T_{\mu\nu}$ and the new source $\theta_{\mu\nu}$ exchange energy and momentum \emph{only gravitationally}: there is no direct microphysical interaction, only interaction mediated by geometry. This is what makes it consistent to speak of $\theta_{\mu\nu}$ as encoding new physics -- a dark sector, an extra fluid, or an effective correction from a modified theory of gravity -- superposed onto a purely GR seed without spoiling the latter's own equilibrium.

Two remarks on the structure of Eqs.~\eqref{eq27}--\eqref{eq29} are in order, since they directly motivate the strategy of Sec.~\ref{sec3}.

\emph{(i) The minimal choice.} From Eq.~\eqref{eq12}, the density depends solely on the radial potential $\mu(r)$. Deforming $\mu(r)$, as in Eq.~\eqref{17}, is therefore a natural and economical choice: it leaves the seed's own density untouched and channels the entire deformation into the pressures and the mass function. The complementary case, deforming the temporal potential instead ($h\neq0$, $f=0$), is a legitimate scheme pursued elsewhere in the literature within the extended (non-minimal) formulation of gravitational decoupling; we do not consider it further here, since our aim is precisely to exploit the minimal case's clean split into an unmodified seed sector and a closed, separately solvable $\theta$-sector.

\emph{(ii) Physical constraints on $f(r)$.} Because $\mu(r)$ is monotonically decreasing on $[0,R]$ for any well-behaved seed, the sign and monotonicity of $f(r)$ are not fixed \emph{a priori}: what matters is that $\mu(r)+\alpha f(r)$ itself remains positive, finite, and monotonic on the same interval, and that $f(0)=0$, so that regularity at the center of the star is preserved. Since $\mu(r)$ controls the seed mass function $m_0(r)=\tfrac{r}{2}[1-\mu(r)]$, the deformation directly modifies the total gravitational mass,
\begin{equation}\label{eq20}
m(r)=\frac{r}{2}\left[1-\mu(r)-\alpha f(r)\right]=m_{0}(r)-\alpha\frac{r}{2}f(r),
\end{equation}
and, at the stellar surface, the compactness factor,
\begin{equation}\label{eq22}
2u(R)=2u_{0}(R)-\alpha f(R), \qquad u_0\equiv M_0/R.
\end{equation}
Thus, depending on the relative sign of $\alpha$ and $f(R)$, gravitational decoupling can either pack additional mass into the same areal radius or lighten the configuration with respect to its isotropic progenitor -- a feature with direct astrophysical relevance for compact stars close to the observational mass ceiling set by pulsars such as PSR J0740+6620 \cite{demorest}.

Equations~\eqref{eq27}--\eqref{eq29} contain four unknowns, $\{\theta^{t}_{t},\theta^{r}_{r},\theta^{\varphi}_{\varphi},f\}$, and only three independent equations (the fourth, Eq.~\eqref{eq30}, being a linear combination of the previous three); the $\theta$-sector is therefore \emph{never} closed by the field equations alone, and extra, physically motivated information must always be supplied by hand. The main strategies pursued in the literature are: (i) \emph{mimic constraints}, in which one of the effective sources is set equal to a component of the seed, e.g. $\theta^{t}_{t}=\bar\rho$ or $\theta^{r}_{r}=\bar p$ \cite{P1,E1,E2,E4,E5,P3}; (ii) direct imposition of a decoupler function $f(r)$ motivated by a known exterior or Weyl-fluid profile \cite{D1,D2,D3}; and (iii) a regularity condition on the $\theta$-sector inspired by the classical Cosenza--Herrera--Esculpi--Witten local anisotropy factor \cite{chew,S1}. Each of these is a well-motivated closure in its own right, built with a specific seed, or a specific physical picture for $\theta_{\mu\nu}$, in mind. Here we pursue a different closure, chosen instead for a mathematical property that none of the above targets: in Sec.~\ref{sec3} we prescribe the \emph{anisotropy-generating} combination $g(r)\equiv\theta^{r}_{r}-\theta^{\varphi}_{\varphi}$ in a functional form such that the resulting linear ODE for $f(r)$ integrates in closed form for the entire two-parameter seed family at once, rather than for one seed at a time.

\section{The Model}\label{sec3}

\subsection{The unified seed space-time}\label{sec3a}

We take as seed an isotropic fluid distribution described by the metric potentials
\begin{equation}\label{nuseed}
\xi(r)=\text{Ln}\left[C\left(1+ar^{2n}\right)^{m}\right],
\end{equation}
\begin{widetext}
\begin{equation}\label{lambdaseed}
\mu(r)=\left(1+ar^{2n}\right)^{2-m}\left(1+a\left(1+mn\right)r^{2n}\right)^{\frac{2m}{1+mn}-2}\left(r^{2}D+F_{1}\!\left(-\tfrac{1}{n};\,1-m;\,\tfrac{2m}{1+mn}-1;\,\tfrac{n-1}{n};\,-ar^{2n};\,-a(1+mn)r^{2n}\right)\right),
\end{equation}
\end{widetext}
where $a$ and $D$ are constants of dimension $\text{length}^{-2}$, and $n,m$ are dimensionless shape parameters. The pair $\{\xi,\mu\}$ solves Eqs.~\eqref{eq23}--\eqref{eq25} identically for any $(n,m)$, by construction, since $\mu(r)$ is obtained by imposing the isotropy condition $8\pi\bar p=$ Eq.~\eqref{eq24} $=$ Eq.~\eqref{eq25} directly on the ansatz~\eqref{nuseed}, following the same logic Durgapal used to generate his own family of solutions \cite{durgapal}.

The interest of the ansatz~\eqref{nuseed} lies in the fact that, for $n=1$, it collapses -- for five particular values of $m$ -- onto some of the best-known and best-behaved isotropic solutions of the Einstein field equations, originally derived independently and by unrelated methods. Table~\ref{tab1} lists these special cases explicitly.

\begin{table}[h]
\caption{Special cases of the unified seed~\eqref{nuseed}--\eqref{lambdaseed} recovered at $n=1$ for different values of $m$.}
\label{tab1}
\begin{tabular}{lcc}
\hline\hline
Model \cite{delgaty} & $n$ & $m$ \\
\hline
Tolman IV \cite{tolman} & $1$ & $1$ \\
Korkina--Orlyanskii \cite{korkina} & $1$ & $2$ \\
Heintzmann IIa \cite{heint} & $1$ & $3$ \\
Durgapal IV \cite{durgapal} & $1$ & $4$ \\
Durgapal V \cite{durgapal} & $1$ & $5$ \\
\hline\hline
\end{tabular}
\end{table}

Each of these five metrics has already been anisotropized via MGD in a dedicated publication, with a closure of the $\theta$-sector specific to that seed. {The construction below unifies the isotropic seeds underlying all of those works under a single prescription, so that their anisotropic extension under the present closure follows from one master formula rather than five separate ones.}

It is worth stressing what changing $n$ physically represents: since $\xi(r)=\text{Ln}\left[C(1+ar^{2n})^{m}\right]$, the parameter $m$ controls the overall ``compactness weight'' of the potential (it is what discriminates between Tolman IV and the stiffer Durgapal V, say, at fixed $n=1$), while $n$ controls how sharply the metric potential departs from flatness as one moves away from the center -- effectively an extra radial-profile handle absent from any of the individually known solutions in Table~\ref{tab1}. 

\subsection{Closing the $\theta$-sector in closed form}\label{sec3b}

To generate an anisotropic compact object out of the seed~\eqref{nuseed}--\eqref{lambdaseed}, we must close the $\theta$-sector, Eqs.~\eqref{eq27}--\eqref{eq29}. Subtracting Eq.~\eqref{eq28} from Eq.~\eqref{eq29} isolates the piece of the $\theta$-sector directly responsible for the local anisotropy induced by MGD
\begin{widetext}
\begin{equation}\label{eq32}
8\pi g(r)= -f\left(\frac{1}{r^{2}}+\frac{\xi^{\prime}}{r}\right)+\frac{f^{\prime}}{4}\left(\xi^{\prime}+\frac{2}{r}\right)
+\frac{f}{4}\left(2\xi^{\prime\prime}+\xi^{\prime2}+2\frac{\xi^{\prime}}{r}\right),
\end{equation}
\end{widetext}
where
\begin{equation}\label{eq33}
g(r)\equiv \theta^{r}_{r}-\theta^{\varphi}_{\varphi}.
\end{equation}
Equation~\eqref{eq32} is a first-order \emph{linear} ODE for $f(r)$, and can therefore always be integrated by quadratures once $g(r)$ (equivalently, the anisotropy one wishes to induce) is prescribed,
\begin{widetext}
\begin{equation}\label{eq34}
\begin{split}
f(r)=r^{2}\left(1+ar^{2n}\right)^{2-m}\left(1+a\left(1+mn\right)r^{2n}\right)^{\frac{2m}{1+mn}-2}\left(B+16\pi\int \frac{g(r)}{r}\left(1+ar^{2n}\right)^{m-1}\left(1+a\left(1+mn\right)r^{2n}\right)^{1-\frac{2m}{1+mn}}dr \right),
\end{split}
\end{equation}
\end{widetext}
with $B$ an integration constant of dimension $\text{length}^{-2}$, fixed below by matching. For a generic choice of $g(r)$, the integral in Eq.~\eqref{eq34} has no closed form for arbitrary $(n,m)$; a closure chosen for one seed will not, in general, remain tractable when carried over to another.

Here we proceed in the opposite direction: we choose $g(r)$ so that the integral in Eq.~\eqref{eq34} is elementary for every $(n,m)$ simultaneously. Choosing
\begin{equation}\label{eq35}
g(r)=\frac{Ar^{2}}{\left(1+ar^{2n}\right)^{m+1} \left(1+a\left(1+mn\right)r^{2n}\right)^{1-\frac{2m}{1+mn}}},
\end{equation}
with $A$ a constant of dimension $\text{length}^{-3}$, the two power-law factors inside the integral of Eq.~\eqref{eq34} cancel exactly against those of $g(r)/r$ except for a single, integrable power of $(1+ar^{2n})$, and one obtains the closed-form master decoupler function
\begin{widetext}
    \begin{equation}\label{eq36}
\begin{split}
f(r)=r^{2}\left(1+ar^{2n}\right)^{2-m}\left(1+a\left(1+mn\right)r^{2n}\right)^{\frac{2m}{1+mn}-2}\left(B+8\pi A\, r^{2}\; {}_2F_{1}\!\left[2,\tfrac{1}{n};\, 1+\tfrac{1}{n};\,-ar^{2n}\right]\right),
\end{split}
\end{equation}
\end{widetext}

valid for \emph{any} real $(n,m)$ for which the seed itself is well defined. This is the central technical result of this work: a single, closed, elementary-function expression for the MGD decoupler function of the entire unified family, which recovers the five classical seeds of Table~\ref{tab1} upon setting $n=1$ and the corresponding value of $m$.

The associated anisotropy factor follows directly from Eqs.~\eqref{eq33} and \eqref{eq35},
\begin{widetext}
\begin{equation}\label{eq31}
\Delta(r;\alpha)\equiv \alpha\left(\theta^{r}_{r}-\theta^{\varphi}_{\varphi}\right)=\alpha\, g(r)=\frac{\alpha A r^{2}}{\left(1+ar^{2n}\right)^{m+1}\left(1+a(1+mn)r^{2n}\right)^{1-\frac{2m}{1+mn}}}.
\end{equation}
\end{widetext}
By construction $\Delta(0)=0$, as required for a regular center, and $\Delta(r)$ is a smooth, single-signed function on $(0,R]$ for suitable choices of $\{a,A,n,m\}$, which is a necessary condition for a physically admissible anisotropic profile (Sec.~\ref{sec5}).

\section{Effective matter content and matching to the exterior}\label{sec4}

Once $f(r)$ is fixed by Eq.~\eqref{eq36}, the full interior geometry is
\begin{equation}\label{eq_geo}
ds^{2}=C\left(1+ar^{2n}\right)^{m}dt^{2}-\left[\mu(r)+\alpha f(r)\right]^{-1}dr^{2}-r^{2}d\Omega^{2},
\end{equation}
with $\mu(r)$ given by Eq.~\eqref{lambdaseed}. The effective, total density and pressures follow from Eqs.~\eqref{eq16}--\eqref{eq18} together with Eqs.~\eqref{eq27}--\eqref{eq29} evaluated on $\xi(r)$:
\begin{eqnarray}
8\pi\rho&=&8\pi\bar\rho -\alpha\left(\frac{f}{r^{2}}+\frac{f^{\prime}}{r}\right), \\
8\pi p_{r}&=&8\pi\bar p +\alpha f\left(\frac{1}{r^{2}}+\frac{\xi^{\prime}}{r}\right), \\
8\pi p_{t}&=&8\pi\bar p +\frac{\alpha f}{4}\left(2\xi^{\prime\prime}+\xi^{\prime2}+2\frac{\xi^{\prime}}{r}\right)+\frac{\alpha f^{\prime}}{4}\left(\xi^{\prime}+\frac{2}{r}\right),
\end{eqnarray}
with $p_t-p_r=\Delta(r;\alpha)$ given by Eq.~\eqref{eq31}. The total mass function is by Eq.~\eqref{eq20}.

At the stellar surface $r=R$, and assuming $\theta_{\mu\nu}=0$ for $r>R$, the interior geometry must match smoothly onto the exterior Schwarzschild vacuum,
\begin{equation}
ds^{2}_{\rm ext}=\left(1-\frac{2M}{r}\right)dt^{2}-\left(1-\frac{2M}{r}\right)^{-1}dr^{2}-r^{2}d\Omega^{2},
\end{equation}
$M$ being the total gravitational mass as measured by a distant observer. The first and second fundamental forms of Darmois and Israel \cite{is,dar} require the continuity of the induced metric and of the extrinsic curvature across $\Sigma:\, r=R$, which for MGD-deformed interiors reduces to
\begin{eqnarray}
\xi(R)&=&\text{Ln}\left(1-\frac{2M}{R}\right), \label{jc1}\\
\mu(R)+\alpha f(R)&=&1-\frac{2M}{R}, \label{jc2}\\
p_{r}(R)&=&\bar p(R)+\frac{\alpha}{8\pi} f(R)\left(\frac{1}{R^{2}}+\frac{\xi^{\prime}(R)}{R}\right)=0. \label{jc3}
\end{eqnarray}
Condition~\eqref{jc3} -- vanishing \emph{total} radial pressure at the boundary, not vanishing $\bar p$ -- is the standard requirement that fixes the stellar radius $R$ once the remaining constants are chosen, and it is here, through the explicit $\alpha f(R)$ term, that the MGD deformation directly feeds back into the determination of the physical size of the star. Equations~\eqref{jc1}--\eqref{jc3}, together with the demand that $\rho, p_r, p_t \geq 0$ and finite everywhere in $[0,R]$, fix the constants $C$, $D$, and $B$ in terms of $M$, $R$, $a$, $A$, $\alpha$, $n$, and $m$.

\section{Physical acceptability criteria}\label{sec5}

\begin{figure*}
\centering
\includegraphics[width=0.95\textwidth]{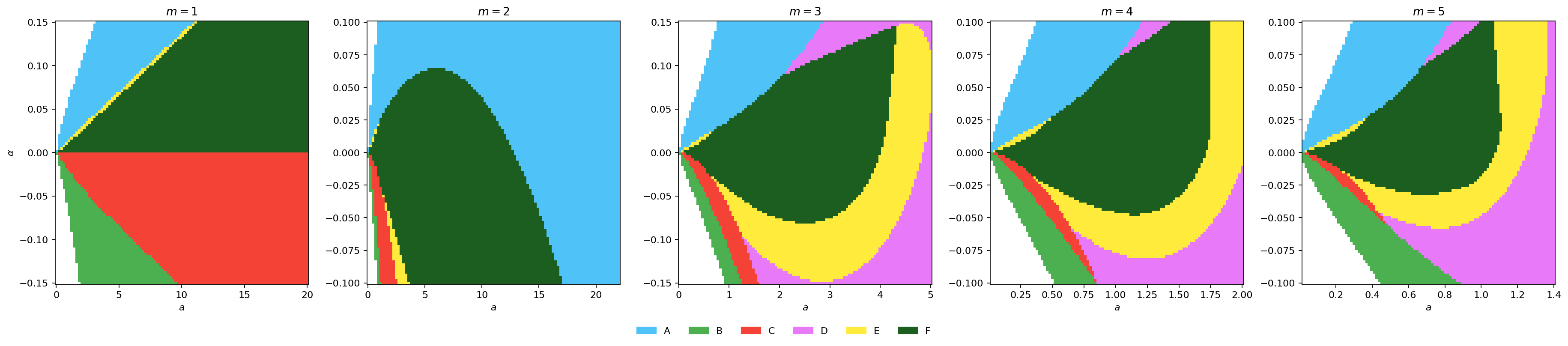}
\caption{Cumulative satisfaction of criteria (i)--(vi) of Sec.~\ref{sec5} over the $(a,\alpha)$ plane, for the five $n=1$ seeds of Table~\ref{tab1} (one panel per model, left to right: Tolman IV, Korkina--Orlyanskii, Heintzmann IIa, Durgapal IV, Durgapal V). Each color marks how many consecutive criteria hold at that point: A (only (i)), B ((i)--(ii)), C ((i)--(iii)), D ((i)--(iv)), E ((i)--(v)), F (all of (i)--(vi)). Only the F region corresponds to a fully admissible configuration. White regions fail even criterion (i). Note the qualitatively different topology of the $F$ region across models: unbounded in $a$ for Tolman IV and Korkina--Orlyanskii, closed off at a finite compactness ceiling for Heintzmann IIa, Durgapal IV, and Durgapal V, as discussed in Sec.~\ref{sec6}.}
\label{fig1}
\end{figure*}

Not every choice of $\{a,A,B,C,D,\alpha,n,m\}$ compatible with Eqs.~\eqref{jc1}--\eqref{jc3} defines a physically admissible stellar model. Following the standard set of criteria used throughout the MGD literature \cite{herre,BOWERS,bohemer,bohemer1,andreasson}, an acceptable solution of the family constructed in Sec.~\ref{sec3}--\ref{sec4} must satisfy, on the whole interior $r\in[0,R]$:

\begin{itemize}
\item[(i)] \textbf{Regularity.} $\xi(r)$, $\mu(r)+\alpha f(r)$, $\rho(r)$, $p_r(r)$, and $p_t(r)$ are finite and free of singularities, with $\Delta(0)=0$ and $e^{\xi(0)}, [\mu(0)+\alpha f(0)]$ finite and non-zero.
\item[(ii)] \textbf{Monotonicity.} $\rho'(r)\leq0$, $p_r'(r)\leq0$, and $p_t'(r)\leq0$ for $r\in(0,R]$, i.e. density and pressures attain their maximum at the center.
\item[(iii)] \textbf{Energy conditions.} $\rho\geq0$, $\rho-p_r\geq0$, $\rho-p_t\geq0$ (dominant energy condition), and $\rho+p_r\geq0$, $\rho+p_t\geq0$, $\rho+p_r+2p_t\geq0$ (null, weak and strong energy conditions).
\item[(iv)] \textbf{Causality.} The radial and tangential sound speeds, $0\leq v_r^{2}=dp_r/d\rho\leq1$ and $0\leq v_t^{2}=dp_t/d\rho\leq1$, must not exceed the speed of light.
\item[(v)] \textbf{Stability against cracking.} The Herrera cracking criterion \cite{herre} requires $-1\leq v_t^{2}-v_r^{2}\leq0$ throughout the interior.
\item[(vi)] \textbf{Adiabatic index.} The relativistic adiabatic index $\Gamma=\frac{\rho+p_r}{p_r}\,\frac{dp_r}{d\rho}$ must satisfy $\Gamma>4/3$ for dynamical stability against radial perturbations, generalized in the anisotropic case by the Bowers--Liang correction \cite{BOWERS}.
\item[(vii)] \textbf{Compactness bound.} The total compactness $2M/R$ must respect the generalized Buchdahl--Andreasson bound for anisotropic matter, $2M/R\leq 8/9$ in the isotropic limit and its anisotropic generalization otherwise \cite{buch,BUCHDAHL150,andreasson}.
\item[(viii)] \textbf{Hydrostatic equilibrium.} The generalized Tolman--Oppenheimer--Volkoff equation,
\begin{equation}
-p_r'-\frac{\xi'}{2}(\rho+p_r)+\frac{2}{r}(p_t-p_r)=0,
\end{equation}
must hold identically; the three contributions are customarily labeled the gravitational ($F_g$), hydrostatic ($F_h$), and anisotropic ($F_a$) forces, whose mutual balance can be visualized explicitly once $(a,A,\alpha,n,m)$ are fixed numerically.
\end{itemize}

\begin{figure*}
\centering
\includegraphics[width=0.98\textwidth]{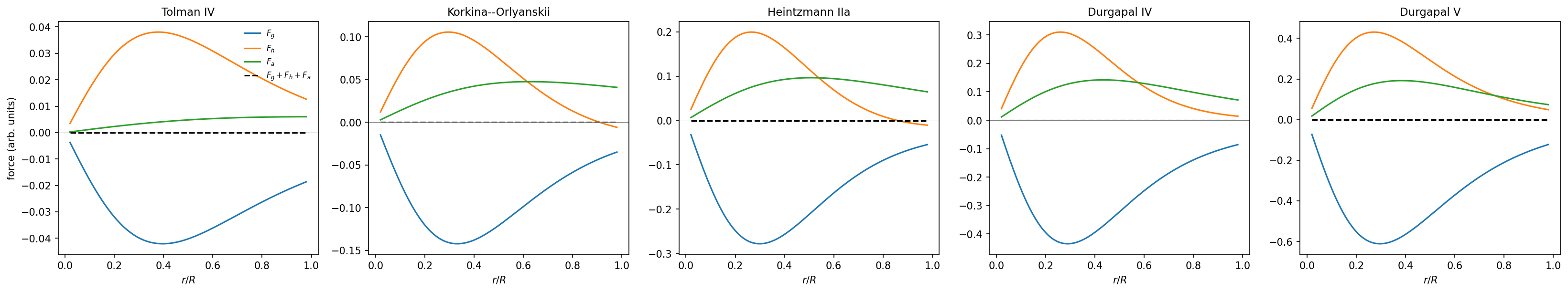}
\caption{{Generalized Tolman--Oppenheimer--Volkoff force balance, criterion (viii), for a representative admissible configuration of each of the five $n=1$ seeds: gravitational ($F_g$), hydrostatic ($F_h$), and anisotropic ($F_a$) forces, and their sum (dashed), which vanishes identically to machine precision throughout the interior for every model.}}
\label{fig3}
\end{figure*}

{Criteria (i)--(vi) are the ones that can fail depending on the choice of $(a,\alpha)$, and Fig.~\ref{fig1} maps exactly where each of them first fails across the $(a,\alpha)$ plane for the five $n=1$ seeds. The region labeled C, where monotonicity is already violated, disappears quickly with growing $a$ for every model; the boundary between D and E marks where the sign of $\Delta$ turns unphysical, while the boundary of the F region itself is set, in every case we examined, by causality (iv) rather than by the energy conditions (iii) or the adiabatic index (vi), which are satisfied on a strictly larger region than F throughout. The qualitative difference already highlighted in the caption -- an $F$ region unbounded in $a$ for $m=1,2$ versus one closed off at a finite compactness ceiling for $m\geq3$ -- is a direct visual signature of the causality obstruction discussed quantitatively in Sec.~\ref{sec6} and Table~\ref{tab2}.
Criteria (vii) and (viii), by contrast, are not choices that can fail within the admissible region: they are consistency checks on the construction itself, and we verify them separately rather than folding them into Fig.~\ref{fig1}. For (vii), the compactness of every configuration in the $F$ region is bounded automatically: since $u(a\to\infty)=m/(m+1)$ at fixed $R$ (Sec.~\ref{sec6}), the seed's own compactness never exceeds $5/6\simeq0.833$ for any $m\leq5$, comfortably below the isotropic Buchdahl bound $8/9$, and we have confirmed numerically that the anisotropic correction $\alpha f(R)$ never pushes a member of the $F$ region past it. For (viii), Fig.~\ref{fig3} shows the explicit force decomposition $F_g$, $F_h$, $F_a$ for a representative admissible configuration of each of the five seeds: the three forces trace markedly different profiles -- $F_g$ is negative and largest in magnitude near $r\simeq0.3R$--$0.4R$, while $F_h$ and $F_a$ are positive and partially compensate it at different radii -- yet their sum vanishes identically to within $10^{-14}$ everywhere in $(0,R)$ for every model, confirming Eq.~\eqref{eq30} numerically on top of its analytic proof and showing that the anisotropic force $F_a$ is never negligible: it is comparable to, and for $r\gtrsim0.5R$ often larger than, the hydrostatic force $F_h$ that would alone balance gravity in the isotropic seed.}

Because $f(r)$ is now known in closed form for arbitrary $(n,m)$, criteria (i)--(viii) can be checked \emph{analytically}, or at worst via a single one-dimensional numerical scan over $\alpha$ at fixed $(n,m)$, rather than through the fully numerical machinery previously required for each seed separately.

\begin{figure*}
\centering
\includegraphics[width=0.85\textwidth]{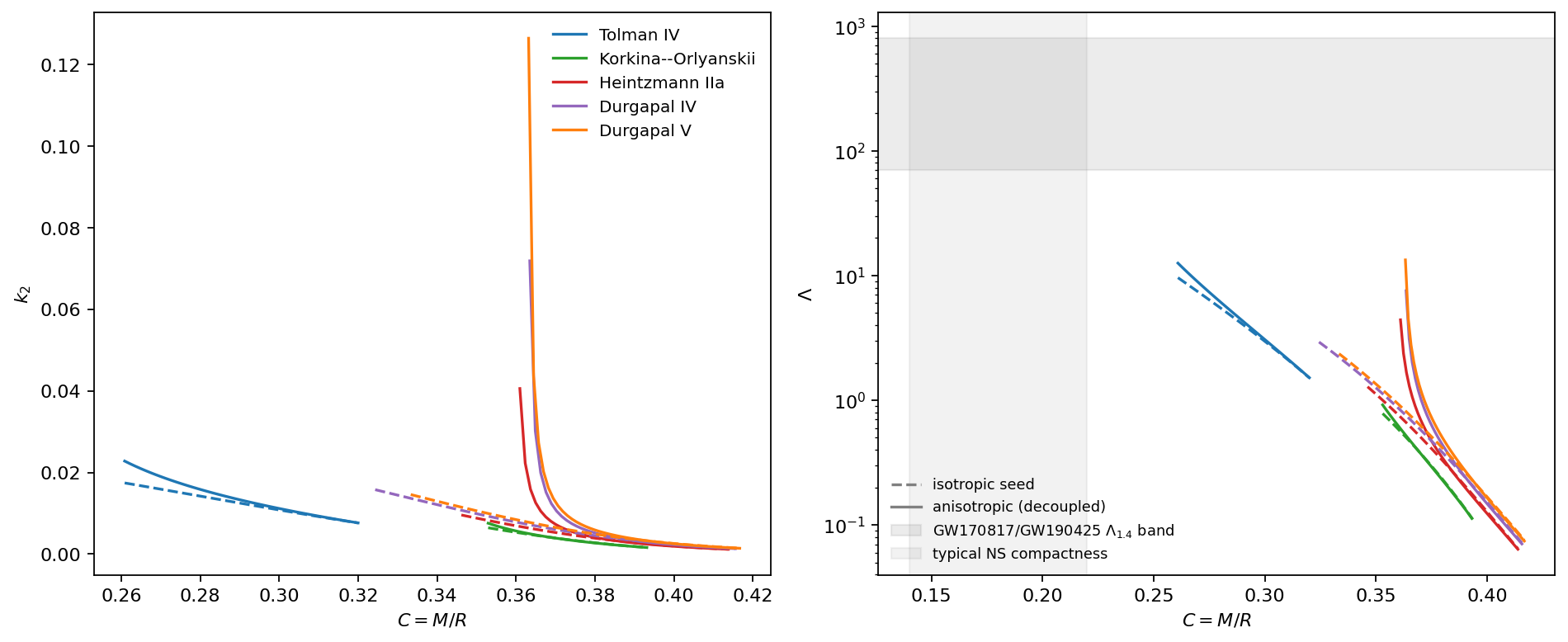}
\caption{Tidal Love number $k_2$ (left) and dimensionless tidal deformability $\Lambda$ (right) as functions of the compactness $C=M/R$, for the five $n=1$ seeds of Table~\ref{tab1}. Dashed: isotropic seed ($\alpha=0$). Solid: anisotropic, decoupled configuration at a representative $\alpha$ inside the admissible window of Table~\ref{tab2}, with $M$ and $R$ held fixed to the seed's own values as $\alpha$ is switched on. For each model, $a$ is scanned over the sub-interval of its admissible range quoted in the text, away from the compactness ceiling discussed below. The shaded horizontal band marks $70\lesssim\Lambda_{1.4}\lesssim800$ from GW170817 \cite{gw170817} (comparable but weaker from GW190425 \cite{gw190425}); the shaded vertical band marks the compactness $C\simeq0.14$--$0.22$ of a typical $1.4\,M_\odot$ neutron star.}
\label{fig2}
\end{figure*}

\section{Astrophysical application}\label{sec6}

The construction above is not merely formal: once $(n,m)$ are fixed to one of the entries of Table~\ref{tab1} and the compact-star data $(M,R)$ of a target configuration are specified, Eqs.~\eqref{jc1}--\eqref{jc3} determine $C$, $D$, and $B$ in terms of $(a,A,\alpha)$, and the resulting interior can be scanned against the acceptability criteria of Sec.~\ref{sec5} (Fig.~\ref{fig1}) to isolate the admissible region of the $(a,\alpha)$ plane. As a direct application, we compute the quadrupolar tidal Love number $k_2$ and the associated dimensionless tidal deformability $\Lambda=\tfrac{2}{3}k_2 C^{-5}$ ($C=M/R$), obtained by solving the static, even-parity perturbation equation for $y(r)=rH'(r)/H(r)$,
\begin{equation}
r y' + y^{2} + y\, e^{\lambda}\!\left[1+4\pi r^{2}(p_r-\rho)\right] + r^{2}Q(r) = 0,
\label{eq_y}
\end{equation}
where, for anisotropic matter, $Q(r)$ is \emph{not} obtained from the isotropic expression by the naive replacement $p\to p_r$: the $\theta\theta$-perturbation equation carries an explicit dependence on how the tangential pressure responds to a radial perturbation, so that
\begin{equation}
Q(r) = 4\pi e^{\lambda}\!\left[5\rho + 9p_r + \frac{\rho+p_r}{A\,c_r^{2}}\right] - \nu'^{2} - \frac{6e^{\lambda}}{r^{2}},
\label{eq_Q}
\end{equation}
with
\begin{equation}
c_r^{2}\equiv\frac{dp_r}{d\rho}, \qquad A\equiv 1+\frac{d\sigma}{dp_r}, \qquad \sigma(r;\alpha)\equiv 2\Delta(r;\alpha),
\label{eq_A}
\end{equation}
$\Delta(r;\alpha)$ being the anisotropy factor of Eq.~\eqref{eq31}, following the generalization of the Hinderer--Damour--Nagar formalism to anisotropic fluids \cite{doneva-yazadjiev}. In the isotropic limit $\Delta\to0$ one has $A\to1$ and Eq.~\eqref{eq_Q} reduces to the standard result. For the present family, both $c_r^{2}$ and $d\sigma/dp_r$ are obtained by differentiating the closed-form $p_r(r)$, $p_t(r)$ of Sec.~\ref{sec4} directly along the radial profile, rather than assumed as an external barotropic input -- so that Eqs.~\eqref{eq_y}--\eqref{eq_A} are fully determined in closed form for arbitrary $(n,m)$, with no need for a numerically inverted equation of state.

We have carried this program out explicitly for the five $n=1$ members of Table~\ref{tab1}, fixing the seed's own surface at $R=1$ through condition~\eqref{jc3} evaluated at $\alpha=0$, and then re-imposing~\eqref{jc3} on the full deformed pressure to fix $B(a,\alpha)$ at the same areal radius; by construction this keeps $M$ and $R$ -- and hence the exterior Schwarzschild spacetime -- exactly fixed as $\alpha$ is varied, isolating the effect of the decoupling on the interior alone. The resulting admissible windows in $\alpha$, at representative values of $a$, are summarized in Table~\ref{tab2}.

\begin{table}[h]
\caption{Representative admissible range of $\alpha$ for each $n=1$ member of Table~\ref{tab1}, at the indicated value of $a$ (units $R=1$, $A=1$), obtained from criteria (i)--(vi) of Sec.~\ref{sec5}. $u\equiv2M/R$ is the compactness of the $\alpha=0$ seed at that $a$.}
\label{tab2}
\begin{tabular}{lccc}
\hline\hline
Model & $a$ & $u$ & admissible $\alpha$ \\
\hline
Tolman IV            & $1$ & $0.500$ & $[\phantom{-}0.000,\ 0.011]$ \\
Korkina--Orlyanskii  & $1$ & $0.667$ & $[-0.10,\phantom{0}0.064]$ \\
Heintzmann IIa       & $1$ & $0.750$ & $[-0.08,\phantom{0}0.144]$ \\
Durgapal IV          & $1$ & $0.800$ & $[-0.048,0.100]$ \\
Durgapal V           & $1$ & $0.833$ & $[-0.031,0.100]$ \\
\hline\hline
\end{tabular}
\end{table}

Two features of Table~\ref{tab2} are worth emphasizing. First, the admissible window narrows monotonically as $m$ grows at fixed $a$: stiffer seeds tolerate less anisotropic deformation before violating causality or the energy conditions. Second, and more strikingly, a full two-dimensional scan of the $(a,\alpha)$ plane shows that the region satisfying \emph{all} of criteria (i)--(vi) simultaneously remains unbounded in $a$ for Tolman IV and Korkina--Orlyanskii (we have verified this up to $a=10^{3}$), while for Heintzmann IIa, Durgapal IV, and Durgapal V it closes off at a finite compactness ceiling ($a\lesssim4.3,\,1.7,\,1.1$, respectively) beyond which \emph{no} choice of $\alpha$ restores causality. This is consistent with the seed's own compactness saturating, $u(a\to\infty)=m/(m+1)$ at $R=1$, at a value that for $m=1,2$ remains safely below both the Buchdahl bound and the causal limit, but that for $m\geq3$ approaches it closely enough that the anisotropic correction can no longer compensate.

With the admissible region in hand, we solve Eq.~\eqref{eq_y} numerically, using the closed-form $\rho(r)$, $p_r(r)$, $p_t(r)$ of Sec.~\ref{sec4} and their exact radial derivatives to evaluate $c_r^{2}$ and $A(r)$ of Eq.~\eqref{eq_A}. Figure~\ref{fig2} shows the resulting $k_2$--$C$ and $\Lambda$--$C$ relations for the five $n=1$ seeds, comparing the isotropic configuration ($\alpha=0$, dashed) with a representative anisotropic member of the admissible window of Table~\ref{tab2} (solid), for a range of $a$ sampled well inside the admissible region.

At fixed compactness, the decoupling systematically raises both $k_2$ and $\Lambda$ relative to the isotropic seed, most visibly for Tolman IV and Korkina--Orlyanskii, and only marginally for Durgapal IV--V, whose admissible $\alpha$-window is an order of magnitude narrower (Table~\ref{tab2}). For Heintzmann IIa, Durgapal IV, and Durgapal V, both $k_2$ and $\Lambda$ turn sharply upward as $a$ approaches the compactness ceiling identified above. We have checked, by re-integrating Eq.~\eqref{eq_y} with three independent solvers (an explicit and two implicit Runge--Kutta schemes) at tolerances down to $10^{-11}$, that $y(R)$ itself varies smoothly through this regime; the upturn in $k_2$ and $\Lambda$ is therefore a genuine feature of the family in this near-critical regime, and not an artifact of the numerical integration, and it appears to be the tidal counterpart of the same causality obstruction that closes the admissible $\alpha$-window in Table~\ref{tab2}.

Figure~\ref{fig2} already places these curves in observational context: the shaded horizontal band marks the $90\%$-credible range $70\lesssim\Lambda_{1.4}\lesssim800$ extracted from GW170817 \cite{gw170817} (GW190425 \cite{gw190425} gives a comparable but less informative bound on $\tilde\Lambda$, owing to the higher component masses of that event), and the shaded vertical band marks the compactness $C\simeq0.14$--$0.22$ typical of a $1.4\,M_\odot$ neutron star with radius $11$--$13$~km. The representative curves of Fig.~\ref{fig2}, computed at $a=1$ purely to illustrate the qualitative effect of the decoupling across the admissible window of Table~\ref{tab2}, sit at compactness $C\gtrsim0.26$ -- well above this band, and correspondingly below it in $\Lambda$. This is a choice of illustration, not a limitation of the family: since $u(a\to\infty)=m/(m+1)$ at fixed $R$, a smaller value of $a$ produces a less compact configuration, and a dedicated scan of $a$ within the admissible window of each model, tuned to land inside the observational compactness band, is needed before a quantitative confrontation with GW170817 and GW190425 is meaningful. What Fig.~\ref{fig2} already establishes is that the qualitative trend -- decoupling raises $\Lambda$ at fixed compactness -- moves the family in the direction that would need to be checked quantitatively against the GW bound once $a$ is tuned to astrophysically relevant compactness.

\section{Conclusions}\label{sec7}

We have shown that five of the best-known isotropic seeds of the gravitational decoupling literature -- Tolman IV \cite{tolman}, Korkina--Orlyanskii \cite{korkina}, Heintzmann IIa \cite{heint}, Durgapal IV \cite{durgapal}, and Durgapal V \cite{durgapal} -- belong to a single two-parameter family generated by $\xi(r)=\text{Ln}[C(1+ar^{2n})^{m}]$, and that this family can be anisotropized via MGD \emph{as a whole}, rather than case by case, by closing the $\theta$-sector with the source function of Eq.~\eqref{eq35}. This choice renders the decoupler function $f(r)$ integrable in closed form, Eq.~\eqref{eq36}, for arbitrary values of the shape parameters $(n,m)$.

{Beyond the closed-form construction itself, mapping the full set of criteria (i)--(viii) across the $(a,\alpha)$ plane for the five $n=1$ members (Fig.~\ref{fig1}) uncovers a structural split within the family: for Tolman IV and Korkina--Orlyanskii the admissible region stays open to arbitrarily high compactness, while for Heintzmann IIa, Durgapal IV, and Durgapal V it is bounded by a compactness ceiling that no choice of the decoupling parameter $\alpha$ can lift, in both cases traced to the causality bound rather than to the energy conditions or stability criteria. The generalized TOV balance (Fig.~\ref{fig3}) confirms that the anisotropic force sourced by $\theta_{\mu\nu}$ is not a small correction to the isotropic equilibrium: across the admissible region it is generally comparable to, and often larger than, the hydrostatic force alone. Using the same closed-form matter content, we computed the quadrupolar tidal Love number $k_2$ and the dimensionless tidal deformability $\Lambda$ across the admissible region for each seed, finding that gravitational decoupling systematically raises both at fixed compactness, with the effect largest for Tolman IV and Korkina--Orlyanskii -- precisely the two branches whose admissible window is widest -- and comparatively muted for Durgapal IV--V, and with a sharp upturn as the compactness ceiling of the Heintzmann and Durgapal branches is approached, which we verified with three independent ODE solvers to be a genuine feature of the near-critical regime rather than a numerical artifact. Placed against the tidal deformability bounds extracted from GW170817 and GW190425, the representative configurations of Fig.~\ref{fig2} sit at higher compactness, and correspondingly lower $\Lambda$, than a canonical $1.4\,M_\odot$ neutron star; the qualitative trend they establish -- decoupling enhances $\Lambda$ at fixed compactness -- is the one that would need to be checked quantitatively once $a$ is tuned, within each model's admissible window, to land inside the observationally relevant compactness range.}

We have laid out the effective matter content, the Darmois--Israel matching conditions to the exterior Schwarzschild vacuum, and the full set of physical acceptability criteria that any admissible member of the family must satisfy, and we have shown how the same closed-form geometry feeds directly into the computation of the tidal Love numbers relevant for gravitational-wave observations of compact-star binaries. {A mass--radius analysis fitted to specific pulsar candidates is the natural next step building on the framework established here.} Charged generalizations of the isotropic ($n=1$) members of the family have also been constructed and are quoted for completeness in the literature; their extension to the present, doubly-parametrized family via MGD is left for future work. 

\section*{acknowledgments}
This work was supported by the Deanship of Scientific Research, Vice Presidency for Graduate Studies and Scientific Research, King Faisal University, Saudi Arabia (Grant No: KFU265375).

\section*{Data availability}
No data were generated or analysed in support of this research. All
results reported are analytic or are reproducible from the equations and
parameter values stated in the text.

\section*{Declaration of Competing Interest}
The authors declare that they have no known competing financial interests or personal relationships that could have appeared to influence the work reported in this paper.

\bibliographystyle{apsrev4-2}
\bibliography{references}

\end{document}